\documentclass[notitlepage,superscriptaddress,nofootinbib,tightenlines,preprintnumbers,twocolumn]{revtex4-2}
\usepackage{amsmath,verbatim,latexsym,amssymb,indentfirst,mathrsfs,mathtools,amsthm,bbm,bm,hyperref,url,cancel,subcaption,enumitem}
\usepackage[font=small,labelfont=bf,format=plain,justification=raggedright,singlelinecheck=false]{caption}
\usepackage{verbatim,indentfirst}
\usepackage[title,titletoc]{appendix}

\usepackage{graphicx}
\usepackage{epstopdf}
\newcommand{\caphead}[1]{{\bf #1}}

\renewcommand{\thesection}{\Roman{section}}
\renewcommand{\thesubsection}{\Roman{section} \Alph{subsection}}
\renewcommand{\thesubsubsection}{\Roman{section} \Alph{subsection} \arabic{subsubsection}}
\makeatletter
\def\p@subsection{}
\makeatother
\makeatletter
\def\p@subsubsection{}
\makeatother

\makeatletter
\newcommand\footnoteref[1]{\protected@xdef\@thefnmark{\ref{#1}}\@footnotemark}
\makeatother

\usepackage{soul}

\usepackage{color}
\usepackage[dvipsnames]{xcolor}
\usepackage[normalem]{ulem}

\newcommand{\Env}{\mathcal{E}}  
\newcommand{\can}{{\rm can}}  
\newcommand{\GC}{{\rm GC}}  
\newcommand{\num}{\mathcal{N}}  
\newcommand{\NATS}{{\rm NATS}}
\newcommand{\avg}{{\rm avg}}
\newcommand{\batt}{\mathcal{B}}  
\newcommand{\MBL}{{\rm MBL}}
\newcommand{\obs}{\mathcal{O}} 
\newcommand{\epo}{\tilde{\Sigma}} 

\newcommand{\hc}{ {\rm h.c.} }

\newcommand{\tot}{ {\rm tot} }
\newcommand{\vN}{{\rm vN}}   
\newcommand{\Tr}{{\rm Tr}}   
\def\id{\mathbbm{1}}   

\newcommand{\Sys}{\mathcal{S}}  
\newcommand{\Sites}{N}  
\newcommand{\var}{{\rm var}}  

\newcommand{\JParen}{ {(j)} }
\newcommand{\KParen}{ {(k)} }
\newcommand{\LParen}{ \bm{(} }
\newcommand{\RParen}{ \bm{)} }

\renewcommand\th{ {\rm th} }

\newcommand*{\bra}[1]{\langle #1\rvert}
\newcommand*{\ket}[1]{\lvert #1 \rangle}

\newcommand*{\ketbra}[2]{\lvert #1 \rangle\!\langle #2 \rvert}
\newcommand*{\expval}[1]{\left\langle  #1  \right\rangle}

\begin{document}
 
\title{Noncommuting conserved charges in quantum thermodynamics and beyond}
\author{Shayan~Majidy} 
\email{smajidy@uwaterloo.ca}
\affiliation{Institute for Quantum Computing, University of Waterloo, Waterloo, Ontario N2L 3G1, Canada}
\affiliation{Perimeter Institute for Theoretical Physics, Waterloo, Ontario N2L 2Y5, Canada} 
\date{\today}
\author{William~F.~Braasch,~Jr.}
\affiliation{Joint Center for Quantum Information and Computer Science, NIST and University of Maryland, College Park, MD 20742, USA}
\author{Aleksander~Lasek}
\affiliation{Joint Center for Quantum Information and Computer Science, NIST and University of Maryland, College Park, MD 20742, USA}
\author{Twesh~Upadhyaya}
\affiliation{Joint Center for Quantum Information and Computer Science, NIST and University of Maryland, College Park, MD 20742, USA}
\affiliation{Department of Physics, University of Maryland, College Park, MD 20742, USA}
\author{Amir~Kalev}
\affiliation{Information Sciences Institute, University of Southern California, Arlington, VA 22203, USA}
\affiliation{Department of Physics and Astronomy, University of Southern California, Los Angeles, California 90089, USA}
\author{Nicole~Yunger~Halpern}
\email{nicoleyh@umd.edu}
\affiliation{Joint Center for Quantum Information and Computer Science, NIST and University of Maryland, College Park, MD 20742, USA}
\affiliation{Institute for Physical Science and Technology, University of Maryland, College Park, MD 20742, USA}

%
%
\begin{abstract}
Thermodynamic systems typically conserve quantities (``charges'') such as energy and particle number. The charges are often assumed implicitly to commute with each other. Yet quantum phenomena such as uncertainty relations rely on observables' failure to commute. How do noncommuting charges affect thermodynamic phenomena? This question, upon arising at the intersection of quantum information theory and thermodynamics, spread recently across many-body physics. Charges' noncommutation has been found to invalidate derivations of the thermal state's form, decrease entropy production, conflict with the eigenstate thermalization hypothesis, and more. This Perspective surveys key results in, opportunities for, and work adjacent to the quantum thermodynamics of noncommuting charges. Open problems include a conceptual puzzle: Evidence suggests that noncommuting charges may hinder thermalization in some ways while enhancing thermalization in others.
\end{abstract}

{\let\newpage\relax\maketitle}

%
%
%
\section{Introduction}
\label{sec_intro}

\begin{figure}[hbt]
\centering
\includegraphics[width=.35\textwidth, clip=true]{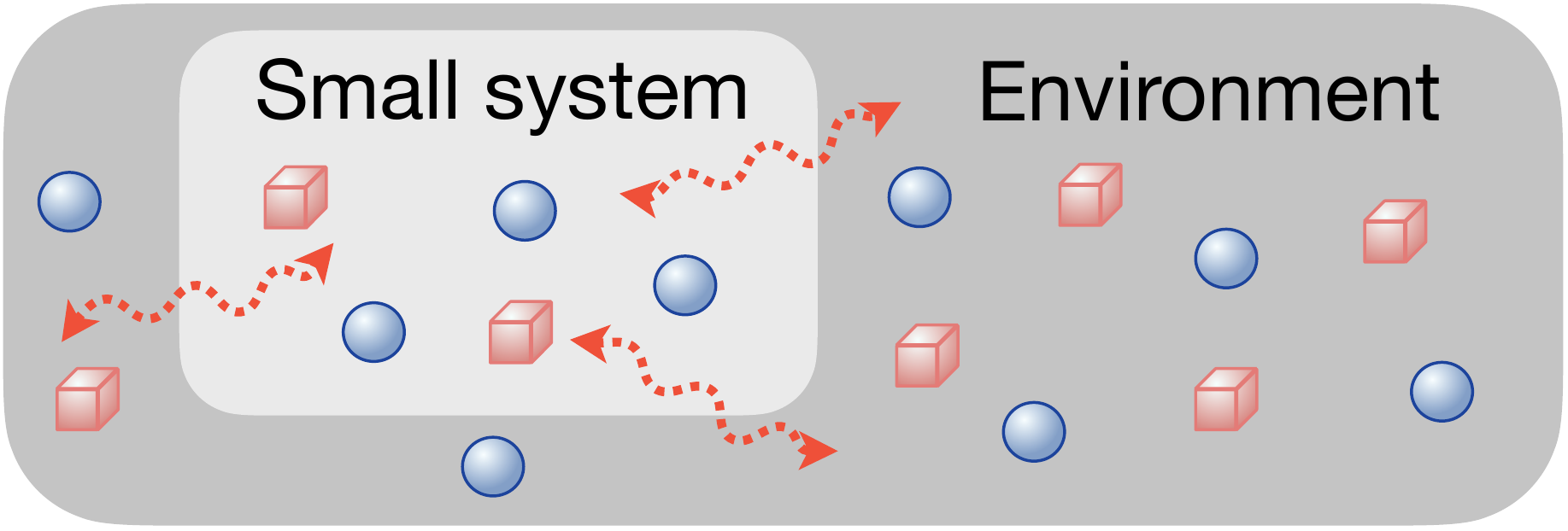}
\caption{\caphead{Common thermodynamic paradigm:} A small system and large environment locally exchange quantities that are conserved globally. Common quantities include energy, as well as particles of different species.}
\label{fig_Basic_Setup}
\end{figure}

A simple story from undergraduate statistical physics motivates this Perspective: Throughout thermodynamics, we consider a small system $\Sys$ exchanging quantities with a large environment $\Env$ (Fig.~\ref{fig_Basic_Setup}). Suppose that the systems are quantum. If they exchange only heat, $\Sys$ may thermalize to the canonical state $\rho_\can  \propto  e^{- \beta H^{(\Sys)} }$. The environment's inverse temperature is $\beta$, and $H^{(\Sys)}$ denotes the system-of-interest Hamiltonian. If $\Sys$ and $\Env$ exchange heat and particles, $\Sys$ may thermalize to the grand canonical state
$\rho_\GC  \propto  e^{- \beta (H^{(\Sys)} - \mu \num^{(\Sys)} ) }$.
The chemical potential is $\mu$, and $\num^{(\Sys)}$ denotes the system-of-interest particle-number operator. This pattern extends to many exchanged quantities (electric charge, magnetization, etc.) and many thermal states.

The exchanged quantities are conserved globally (across $\Sys \Env$), so we call them \emph{charges}. Hermitian operators $Q_a$ represent the conserved quantities; $\Sys$ has an operator $Q_a^{(\Sys)}$, $\Env$ has $Q_a^{(\Env)}$, and the global system has
$Q_a^\tot  \coloneqq  Q_a^{(\Sys)} + Q_a^{(\Env)}  
\equiv  Q_a^{(\Sys)} \otimes \id^{(\Env)}  
           +  \id^{(\Sys)}  \otimes  Q_a^{(\Env)}$. 
The index $a = 0, 1, \ldots, c$; and the Hamiltonian $H = Q_0$.

Across thermodynamics, we often assume implicitly that the charges commute with each other: 
$[Q_a, Q_{a'}] = 0$ $\forall a, a'$. This assumption is almost never mentioned (recall your undergraduate statistical-physics course). However, the assumption underlies derivations of the form of thermal states~\cite{NYH_18_Beyond,NYH_16_Microcanonical}, linear-response coefficients~\cite{Manzano_22_Non}, and more. Observables' ability to fail to commute, though, enables quintessentially quantum phenomena: the Einstein--Podolsky--Rosen paradox~\cite{Einstein_35_Can}, uncertainty relations~\cite{Coles_17_Entropic}, measurement disturbance~\cite{Benitez_19_Survey}, etc. Quantum physics therefore compels us to ask, \emph{what happens to thermodynamic phenomena under dynamics that conserve charges $Q_a^\tot$ that fail to commute with each other?}

This question arose recently at the intersection of quantum information theory and quantum thermodynamics, then seeped into many-body physics. This Perspective surveys results, opportunities, and adjacent work. In the rest of the introduction, we first present a simple physical example. Then, we sample example phenomena transformed by charges' noncommutation. We later establish notation, followed by the Perspective's outline.

\begin{figure}[hbt]
\centering
\includegraphics[width=.35\textwidth, clip=true]{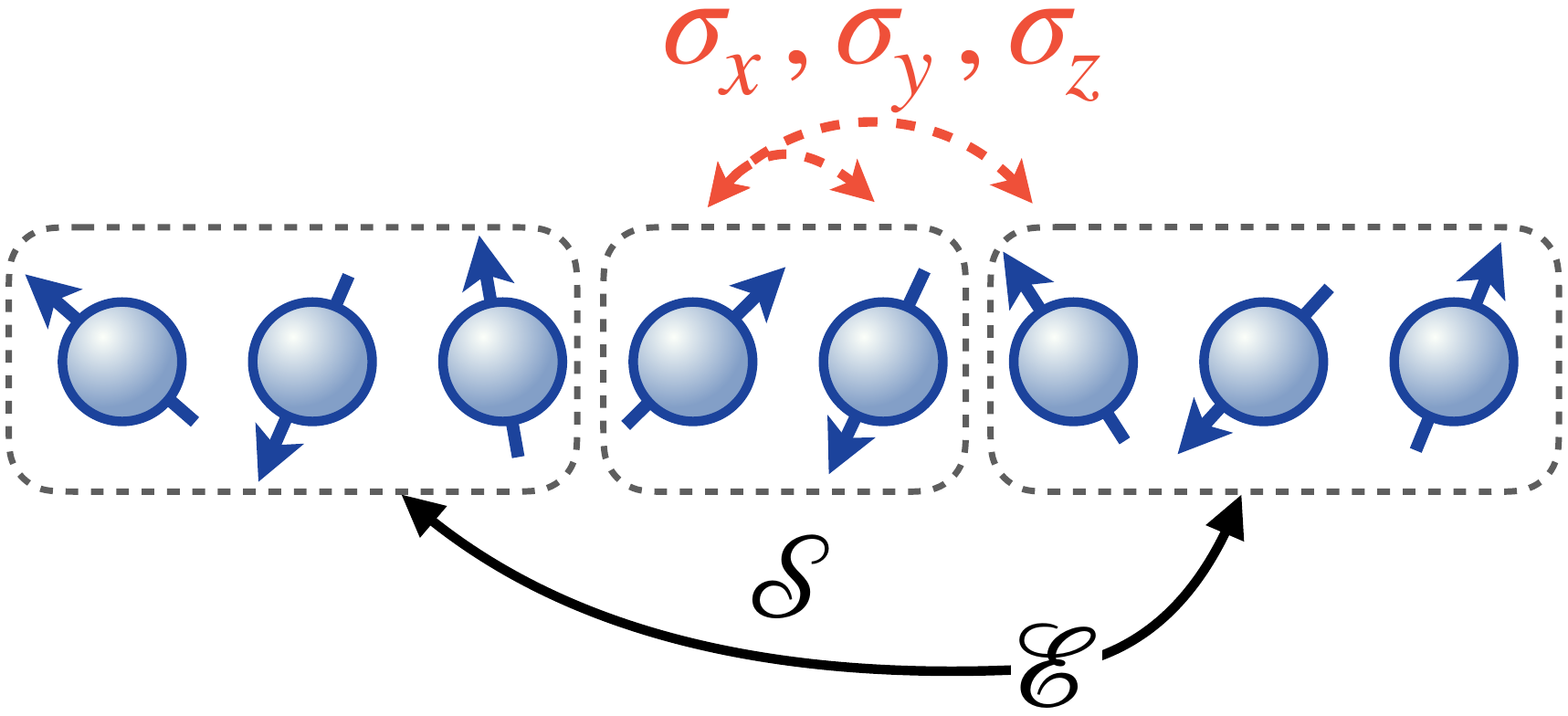}
\caption{\caphead{Example thermodynamic system that conserves noncommuting charges:} 
Two qubits form the system $\Sys$ of interest, and the rest form the environment $\Env$. A qubit's three spin components, $\sigma_{a = x,y,z}$, form the local noncommuting charges. The dynamics locally transport and globally conserve the charges.
}
\label{fig_Qubit_Chain}
\end{figure}

A simple physical example was proposed theoretically~\cite{NYH_20_Noncommuting,Majidy_23_Non} and realized with trapped ions~\cite{Kranzl_23_Experimental}: a chain, shown in Fig.~\ref{fig_Qubit_Chain}, formed from qubits (quantum two-level systems). A few (e.g., two) qubits form $\Sys$, and the other qubits form $\Env$. The chain constitutes a closed quantum many-body system, of the sort whose internal thermalization has recently been studied theoretically and experimentally (e.g.,~\cite{Kaufman_16_Quantum, Neill_16_Ergodic, Clos_16_Thermalization, Zhou_22_Thermalization}). Denote by 
$\sigma_a$ the Pauli-$a$ operator, for $a = x, y, z$; by
$\sigma_a^\JParen$, a spin component of qubit $j$; and by
$\sigma_a^\tot  \coloneqq  \sum_j  \sigma_a^\JParen$,
a total spin component.
One can construct a Hamiltonian that overtly transports quanta of each 
$\sigma_a$ locally while conserving the 
$\sigma_a^\tot$'s globally~\cite{NYH_20_Noncommuting,Majidy_23_Non}:
Denote the $\sigma_z$ ladder operators by 
$\sigma_{\pm z}  \coloneqq  \frac{1}{2} ( \sigma_x  \pm  i \sigma_y )$.
The operator $\sigma_{+z}^\JParen  \sigma_{-z}^{(j+1)}  +  \hc$ transports
one $\sigma_z$ quantum from qubit $j+1$ to qubit $j$ and vice versa, in superposition.
Define ladder operators and couplings analogously for $\sigma_x$ and $\sigma_y$.
The Hamiltonian 
\begin{align}
   \label{eq_Heis}
   H_{\rm Heis}^\tot  =  \sum_j  \sum_{\alpha = x,y,z}  
   \left( \sigma_{+\alpha}^\JParen  \sigma_{-\alpha}^{(j+1)}  +  \hc \right)
   =  \sum_j  \vec{\sigma}^\JParen  \cdot  \vec{\sigma}^{ (j+1) } 
\end{align}
transports the $\sigma_a$'s locally, while conserving them globally.
$H_{\rm Heis}^\tot$ is the Heisenberg model, theoretically well-known and experimentally realizable~\cite{NYH_20_Noncommuting,Kranzl_23_Experimental}.
Yet $H_{\rm Heis}^\tot$ is rarely, if ever, expressed as in~\eqref{eq_Heis} outside the thermodynamics of noncommuting charges---as locally transporting and globally conserving three noncommuting charges. Furthermore, we can extend $H_{\rm Heis}^\tot$ to nonintegrable models (which promote thermalization), to subsystems beyond qubits, and to charges beyond spin components~\cite{Majidy_23_Non}.\footnote{
For more quantum thermodynamics of spin exchanges, see~\cite{Vaccaro_11_Information,Croucher_17_Discrete,Wright_18_Quantum,Croucher_21_Memory}.
}

Having exhibited a thermodynamic system whose $Q_a$'s fail to commute, we sample the physics altered by charges' noncommutation. This sample only dips into the known results (detailed in Sec.~\ref{sec_new_phys}) but demonstrates why noncommuting thermodynamic charges merit study. If the $Q_a$'s fail to commute, how much $\Sys$ can thermalize is unclear, for several reasons. First, noncommuting charges impede two derivations of the thermal state's form~\cite{NYH_18_Beyond,NYH_16_Microcanonical}. 
Second, noncommuting charges force degeneracies on $H^\tot$. 
The reason is Schur's lemma, a group-theoretic result (App.~\ref{sec_Schur} and~\cite{Bouchard_20_Group,Gour_08_Resource,NYH_20_Noncommuting}). 
Nondegenerate Hamiltonians underlie arguments supporting thermalization, mixing, and equilibration (e.g.,~\cite{Deutsch_91_Quantum,Srednicki_94_Chaos,deOliveira_18_Equilibration,Chakraborty_20_How})---arguments challenged by noncommuting charges.

Third, noncommuting charges conflict with the eigenstate thermalization hypothesis (ETH)~\cite{Murthy_23_Non}. The ETH is a theoretical and conceptual toolkit for explaining how quantum many-body systems thermalize internally~\cite{Deutsch_91_Quantum,Srednicki_94_Chaos,Rigol_08_Thermalization}. Myriad numerical and experimental observations support the ETH, which has been applied across many-body physics. Yet noncommuting charges preclude the ETH (Sec.~\ref{sec_new_phys}). Hence noncommuting charges challenge expectations about thermalization in a third manner. 

We will encounter even more manners across this Perspective. Resistance to thermalization matters fundamentally and for applications: Most systems thermalize, so resistance is unusual. Few mechanisms for resistance in quantum many-body systems are known~\cite{Abanin_19_Colloquium,Regnault_22_Quantum}. 
Additionally, thermalization disperses information about initial conditions, so thermalization resistance can promote memory storage.

A major research opportunity is to identify to what extent noncommuting charges hinder thermalization. Another opportunity is to re-examine every thermodynamic or chaotic phenomenon, asking to what extent it changes if charges fail to commute. Phenomena known to change, as reviewed in this Perspective, include the ETH, microcanonical states, thermodynamic-entropy production, entanglement entropy, and local interactions' ability to effect global evolutions. A third opportunity is to marry recent noncommuting-charge work, based largely in quantum thermodynamics and information theory, with adjacent work in other fields, such as high-energy theory. Other fields favor the terminology \emph{non-Abelian symmetries}; fundamental thermodynamics motivates our noncommuting-charge language. A Hamiltonian conserves noncommuting charges, 
$[H^\tot, Q_a^\tot] = 0$ $\forall a$, if and only if it has a continuous non-Abelian symmetry: 
$U_a^\dag H^\tot U_a = H^\tot$ $\forall a$. 
The charges generate the unitary operators: $U_a = e^{-i Q_a t}$, for $t \in \mathbb{R}$. 
The $Q_a$'s form a Lie algebra, and the $U_a$'s form the associated Lie group~\cite{NYH_22_How}.
We flesh out the three above-mentioned opportunities across this Perspective.

Across the Perspective, we use the notation introduced above, as well as the following. We call a closed quantum many-body system (e.g., the composite $\Sys \Env$ in Fig.~\ref{fig_Basic_Setup}) a \emph{global} system. $\Sites$ denotes the number of degrees of freedom (DOFs) in a global system. Often, $\Sys \Env$ will consist of $\Sites$ copies of $\Sys$. For example, $\Sites$ denotes the number of qubits in Fig.~\ref{fig_Qubit_Chain}. Large but finite $\Sites$---the mesoscale---interests us: As $\Sites \to \infty$, $\Sys \Env$ grows classical, according to the correspondence principle~\cite{NYH_16_Microcanonical}. Noncommutation enables nonclassical phenomena, so we should expect charges' noncommutation to influence thermodynamic phenomena at finite $\Sites$.
Continuing with notation, we ascribe to the Pauli operators $\sigma_a$ eigenstates $\ket{ a \pm }$ associated with the eigenvalues $\pm 1$. Subscripts index charges (as in $\sigma_a^\JParen$), whereas superscripts index sites or other subsystems (e.g., $\Sys$ and $\Env$). We often denote commuting charges, or other observables that might commute, by $\tilde{Q}_a$'s.

The rest of this Perspective is organized as follows. Section~\ref{sec_early_work} introduces the earliest work on noncommuting thermodynamic charges. Section~\ref{sec_new_phys} surveys new physics that arises from charges' noncommutation. In Sec.~\ref{sec_adjacent}, we discuss bridges with related research. Section~\ref{sec_outlook} crystallizes the richest-looking avenues for future work.

\section{Early work}
\label{sec_early_work}

Jaynes was the first, to our knowledge, to address anything like noncommuting thermodynamic charges~\cite{Jaynes_57_Info_II}. 
He formalized the \emph{principle of maximum entropy}, which pinpoints the state 
$\rho = \sum_k p_k \ketbra{k}{k}$ 
most reasonably attributable to a system about which one knows little. Imagine knowing about $\rho$ only the expectation values 
$\langle \tilde{Q}_a \rangle$ of observables 
$\tilde{Q}_a$. 
The state obeys the constraints 
$\Tr( \rho \tilde{Q}_a ) = \tilde{Q}_a^\avg \in \mathbb{R}$,
plus the normalization condition
$\Tr (\rho) = 1$.
Whichever constraint-obeying state maximizes the von Neumann entropy $S_\vN (\rho) \coloneqq 
- \Tr \LParen \rho \log(\rho) \RParen$
is most reasonable,
according to the maximum-entropy principle.
(This subsection's logarithms are base-$e$.)
The entropy maximization encapsulates our ignorance of everything except the constraints. The function maximized is
\begin{align}
   \label{eq_Lagrangian}
   & S_\vN(\rho) - 0  \\
   & = S_\vN(\rho) - \lambda \left[ \Tr(\rho) - 1 \right]
   - \sum_a \tilde{\mu}_a \left[ 
     \Tr(\rho \tilde{Q}_a) - \tilde{Q}_a^\avg \right] 
   \nonumber \\ \nonumber
   & \eqqcolon \mathcal{L}(\rho, \lambda, \{ \tilde{\mu}_a \}) .
\end{align}
Lagrange multipliers are denoted by $\lambda, \tilde{\mu}_a \in \mathbb{R}$.
The state's eigenbasis, $\{ \ketbra{k}{k} \}$, equals an eigenbasis of $\sum_a \tilde{\mu}_a \tilde{Q}_a$:
If $\{ \ketbra{k}{k} \}$ did not, decohering $\rho$ with respect to any 
$\sum_a \tilde{\mu}_a \tilde{Q}_a$ eigenbasis would raise
$S_\vN$ monotonically, while preserving the other terms in~\eqref{eq_Lagrangian}~\cite{Guryanova_16_Thermodynamics}. Hence the decohered state would achieve at least as large an $\mathcal{L}$ value as $\rho$.
Maximizing $\mathcal{L}(\rho, \lambda, \{ \tilde{\mu}_a \})$ with respect to $p_k$ yields
\begin{align}
   \label{eq_GGE}
   \rho 
   = \frac{1}{Z} \,
   e^{- \sum_a \tilde{\mu}_a \tilde{Q}_a } \, .
\end{align}
Maximizing with respect to $\lambda$ fixes the partition function: 
$Z = \Tr( e^{- \sum_a \tilde{\mu}_a \tilde{Q}_a } )$.
Maximizing with respect to $\tilde{\mu}_a$ yields
$\tilde{Q}_a^\avg = - \frac{\partial}{\partial \tilde{\mu}_a } \log(Z)$.
This procedure works, Jaynes noted, even if the $\tilde{Q}_a$'s do not commute. He showed how to measure such observables' $\tilde{Q}_a^\avg$'s using ancillas.
One can derive his results alternatively via analytic properties of $Z$~\cite{Liu_06_Gibbs}.

Jaynes's work, though pioneering, left threads hanging. First, the maximum-entropy principle invokes only information theory and quantum physics, not thermodynamics or charge conservation. Connecting $\rho$ to equilibrium requires more-physical arguments. Second, Jaynes wrote only one paragraph about noncommuting $\tilde{Q}_a$'s. What could they represent physically? Which systems might have them? When might they impact thermodynamics? 

Still, Jaynes's known-about observables $\tilde{Q}_a$ can be noncommuting charges $Q_a$. Hence Jaynes was the first to write what was later termed the \emph{non-Abelian thermal state} (NATS)~\cite{NYH_16_Microcanonical}, 
\begin{align}
   \label{eq_NATS}
   \rho_\NATS
   \coloneqq \frac{1}{Z} e^{-\beta (H - \sum_a \mu_a Q_a)}.
\end{align}
$\beta$ denotes an inverse temperature.
The $\mu_a$'s denote generalized chemical potentials.
$\rho_\NATS$ has the form of Eq.~\eqref{eq_GGE}, which has been called the generalized Gibbs ensemble (GGE), regardless of whether the charges commute~\cite{Rigol_07_Relaxation, Rigol_09_Breakdown, Langen_15_Experimental, Vidmar_16_Generalized}. We use the term NATS for three reasons. First, the term GGE was initially introduced for integrable systems, whereas the NATS is a thermal state. 
Second, most GGE literature concerns commuting charges (see Sec.~\ref{sec_adjacent} for exceptions). Third, justifying the NATS's form physically (not merely information-theoretically) is more difficult than justifying the form of a typical (commuting-charge) GGE, as explained later in this section.

Acolytes of Jaynes's expanded upon his work ~\cite{Balian_86_Dissipation,Elze_86_Quantum,Alhassid_78_Connection,Balian_87_Equiprobability}.
Balian and Balazs ~\cite{Balian_87_Equiprobability} sought a more-physical justification for $\rho_\NATS$'s form. They imagined $\Sites$ copies of the system of interest, in the ensemble tradition of thermodynamics. In thermodynamics, we regard all copies except one ($\Sys$) as forming an effective environment ($\Env$)~\cite{Callen_85_Thermodynamics}.  
Imagine $\Sys$ exchanging energy and particles with $\Env$. How can we prove that $\Sys$ is in a canonical state $\rho_\can$? We assume that $\Sys \Env$ has a fixed particle number and an energy in a small window---that
$\Sys \Env$ is in a microcanonical subspace. Tracing out $\Env$ from the microcanonical state yields $\Sys$'s state, which equals $\rho_\can$ if $\Sys$ and $\Env$ couple weakly~\cite{Laundau_80_Statistical}. Suppose that $\Sys$ and $\Env$ exchange several commuting charges $\tilde{Q}_a$. The microcanonical subspace is an eigenspace shared by the $\tilde{Q}_a^\tot$'s (except $\tilde{Q}_0^\tot = H^\tot$, the time-evolution generator). What if $\Sys$ and $\Env$ exchange noncommuting charges $Q_a$? The $Q_a^\tot$'s share no eigenbasis, so they might share no eigenspaces. Microcanonical subspaces might not exist.

Balian and Balazs tried to overcome this challenge. 
They observed that the charge densities 
$Q_a^\tot / \Sites$ commute in 
the infinite-$\Sites$ limit:
$\lim_{\Sites \to \infty} \frac{1}{\Sites^2}
[Q_a^\tot, Q_{a'}^\tot] = 0$ $\forall a, a'$.
But they could not construct a well-justified generalization of microcanonical subspaces for noncommuting charges. That goal would wait for the maturation of quantum information theory.

For decades afterward, no literature addressed thermodynamic charges' ability to not commute, to our knowledge. The topic gained attention a few years ago at the intersection of quantum information and quantum thermodynamics. Two publications showed that noncommuting charges can violate thermodynamic expectations~\cite{Lostaglio_14_Thermodynamics,NYH_18_Beyond}. 

Lostaglio demonstrated that noncommuting charges overturn an expectation about free energy~\cite{Lostaglio_14_Thermodynamics}. 
Consider a system $\Sys$ of interest in a state $\rho^{(\Sys)}$. Let $\Sys$ begin uncorrelated with its environment $\Env$, which consists of $c$ subsystems: 
$\rho^{(\Sys)} \otimes \rho^{(\Env)}$, wherein
$\rho^{(\Env)}$ has the GGE form~\eqref{eq_GGE}.
$\Sys$ has $c$ commuting charges $\tilde{Q}_a^{(\Sys)}$, 
and $\Env$ has $\tilde{Q}_a^{(\Env)}$'s. 
One can attribute to $\Sys$ a ``free energy''\footnote{
The quotation marks reflect the controversy surrounding free energies defined information-theoretically, for out-of-equilibrium states.}
$F_a^{(\Sys)} (\rho) \coloneqq 
- \frac{1}{\beta \mu_a} S(\rho) 
+ \Tr (\tilde{Q}_a \rho)$.
Let any charge-conserving unitary $U$ evolve $\Sys \Env$ to a state $\rho^{(\Sys \Env)}_{\rm f}$:
$[U, \tilde{Q}_a^\tot] = 0$ $\forall a$.
Subscript ${\rm f}$'s will distinguish $\Sys$'s and $\Env$'s final states.
$\Env$'s $j^\th$ subsystem ends up in
$\rho^{(\Env, j)}_{\rm f} 
\coloneqq \Tr_{ \bar{j} } ( \rho^{(\Env)}_{\rm f} )$.
Three more quantities change:
$\Sys$'s von Neumann entropy, by
$\Delta S^{(\Sys)}_\vN 
\coloneqq S_\vN ( \rho^{(\Sys)}_{\rm f} ) 
- S_\vN ( \rho^{(\Sys)} )$;
the $a^\th$ charge's environmental expectation value, by 
$\Delta  \langle \tilde{Q}_a^{(\Env)}  \rangle 
\coloneqq \mathrm{Tr} ( \tilde{Q}_a^{(\Env)} [\rho^{(\Env)}_{\rm f} - \rho^{(\Env)}] )$;
and the system's $a^\th$ ``free energy,'' by $\Delta F_a^{(\Sys)}$.
$U$ redistributes charges and information contents according to
\begin{align}
   \label{eq_matteo_thesis}
   & - \Delta S^{(\Sys)}_\vN 
   + D \left( \rho^{(\Sys \Env)}_{\rm f} 
   \left\lvert \right\rvert
   \rho^{(\Sys)}_{\rm f} \otimes \rho^{(\Env)}_{\rm f} \right)  
   \\ \nonumber \quad
   & = \beta \sum_a  \mu_a 
   \left( \Delta  \langle \tilde{Q}_a^{(\Env)}  \rangle
          - \Delta F_a^{(\Sys)} \right) 
   - D \left( \rho^{(\Env)}_{\rm f} 
              \left\lvert \right\rvert
              \otimes_j \rho^{(\Env, j)}_{\rm f} \right). 
\end{align}
The relative entropy 
$D(\sigma_1 || \sigma_2) 
= -\Tr(\sigma_1 [\log \sigma_1 - \log \sigma_2] )$ quantifies the distance between quantum states $\sigma_{1,2}$~\cite{NielsenC10}.
Equation~\eqref{eq_matteo_thesis} relates that the growth in $\Sys$'s information content, plus the global state's final nonseparability, depends on the average changes in $\Env$'s charges, $\Sys$'s ``free-energy'' change, and the correlations formed in $\Env$.
If charges fail to commute, the derivation breaks down. No term decomposes into distinct charges' $\Delta F_a^{(\Sys)}$.
Hence noncommuting charges undermine an expectation about free energy.

Second, Yunger Halpern reasoned about noncommuting charges in thermodynamic resource theories~\cite{NYH_18_Beyond}. 
\emph{Resource theories} are information-theoretic models for contexts in which restrictions constrain the operations performable and the systems accessible~\cite{Chitambar_19_Quantum}. Using a resource theory, one calculates the optimal efficiencies with which an agent, subject to the restrictions, can perform tasks such as extracting work from nonequilibrium quantum systems.
In thermodynamics, the first law constrains agents to conserve energy. Every unitary $U$ performable on a closed, isolated system conserves the total Hamiltonian: 
$[U, H^\tot] = 0$~\cite{Lostaglio_19_Introductory}.
Suppose that $U$ must conserve commuting global charges $\tilde{Q}_a^\tot$: 
$[U, \tilde{Q}_a^\tot] = 0$~\cite{NYH_16_Beyond}.
Which systems, if freely accessible to a thermodynamic agent, would render the model nontrivial---would not enable the agent to, say, perform work for free, achieving a \emph{perpetuum mobile}?
Systems in the equilibrium state~\eqref{eq_GGE}~\cite{NYH_16_Beyond,Brando_15_Second}.
The proof fails, however, if the charges fail to commute~\cite{NYH_18_Beyond}.

References~\cite{Lostaglio_14_Thermodynamics,NYH_18_Beyond} spurred three groups to apply quantum-information-theoretic thermodynamics to noncommuting charges~\cite{Guryanova_16_Thermodynamics, Lostaglio_17_Thermodynamic,NYH_16_Microcanonical}
(as overviewed in~\cite{Hinds_18_Quantum}).
Guryanova \textit{et al.} quantified tradeoffs among work and charges, using a similar resource theory~\cite{Guryanova_16_Thermodynamics}. Consider a system $\Sys$ in an out-of-equilibrium state $\rho^{(\Sys)}$. $\Sys$ begins uncorrelated with an environment $\Env$ in a GGE $\rho^{(\Env)}$. 
Each of $\Sys$ and $\Env$ has charges $Q_a$ that might or might not commute. 
(In this section, only in this paragraph do $Q_a$'s denote possibly commuting charges.) The setup includes also \emph{batteries}, systems in which work or charges can be reliably stored and from which the resources can be reliably retrieved. 
The authors ascribe a \emph{free entropy} to every state $\rho$ of $\Sys$:
\begin{equation}
    \tilde{F}^{(\Sys)} (\rho)
    \coloneqq \beta \sum_a \mu_a 
    \Tr \left( \rho Q^{(\Sys)}_a \right) 
    - S_\vN \left( \rho^{(\Sys)} \right).
\end{equation}
\{$\tilde{F}^{(\Sys)} (\rho)$ is the negative of a nonequilibrium extension of a Massieu function~\cite{Callen_85_Thermodynamics}.\}
Let $\Sys\Env$ evolve under any unitary $U$ that conserves each total (system--environment--battery) charge.
For all $X = \Sys, \Env$, 
the state $\rho^{(X)}$ evolves to $\rho^{(X)}_{\rm f}$.
$X$'s average $a$-charge changes by 
$\Delta \langle Q_a^{(X)}  \rangle
\coloneqq \Tr (Q_a^{(X)} [ \rho^{(X)}_{\rm f} - \rho^{(X)} ] )$. 
$\Sys$'s free entropy changes by
$\Delta \tilde{F}^{(\Sys)}$.
The batteries perform on $\Sys \Env$ the average  ``charge-$a$ work'' 
$W_a = - \big( \Delta \langle Q_a^{(\Sys)} \rangle 
+ \Delta \langle Q_a^{(\Env)}  \rangle  \big)$.
This charge work obeys a generalized second law of thermodynamics:
\begin{align}
\label{eq:secondlaw}
    \beta \sum_a \mu_a  W_a 
    \leq -\Delta \tilde{F}^{(\Sys)}.
\end{align}
If $\Sys$ is trivial (corresponds to a 0-dimensional Hilbert space) or is unchanged by $U$,
$\Delta \tilde{F}^{(\Sys)} = 0$. Consequently, one can extract any amount of any charge from $\Env$, by paying a price in the other charges: 
$\mu_{a}  W_{a} \leq -\sum_{a' \neq a} \mu_{a'}  W_{a'}$.
Under what conditions Eq.~\eqref{eq:secondlaw} is saturable depends on whether the charges commute.
Lostaglio \emph{et al.}, too, drew tradeoff conclusions, which they applied to the erasure of qubits~\cite{Lostaglio_17_Thermodynamic}. 

Rounding out the trio~\cite{Guryanova_16_Thermodynamics, Lostaglio_17_Thermodynamic,NYH_16_Microcanonical},
Yunger Halpern \emph{et al.} physically justified $\rho_\NATS$'s form in multiple ways, beginning with a microcanonical-like derivation~\cite{NYH_16_Microcanonical}. 
Independently of Balian and Balazs~\cite{Balian_87_Equiprobability}, the authors realized that noncommuting charges prevent microcanonical subspaces from existing (in abundance). 
The authors therefore generalized microcanonical to \emph{approximate microcanonical} (AMC) subspaces. In an AMC subspace $\mathcal{M}$, every $Q_a^\tot$ has a fairly well-defined value: Measuring any $Q_a^\tot$ has a high probability of yielding a value near the expectation value $\langle Q_a^\tot \rangle$. The authors defined $\mathcal{M}$ and proved its existence under certain conditions~\cite{NYH_16_Microcanonical}.
Denote by $\Pi_{\mathcal{M}}$ the projector onto $\mathcal{M}$. Consider ascribing the AMC state 
$\Pi_{\mathcal{M}} / \Tr( \Pi_{\mathcal{M}} )$ to the global system, formed from $\Sites$ copies of the system $\Sys$ of interest. Trace out all copies except the $\ell^\th$: 
$\rho^{(\ell)} = \Tr_{ \bar{\ell} } 
\LParen \Pi_{\mathcal{M}} / \Tr( \Pi_{\mathcal{M}} ) \RParen$. Compare $\rho^{(\ell)}$ with $\rho_\NATS$ using the relative entropy. Average over $\ell$. This average distance is upper-bounded as
\begin{align}
   \big\langle D \big( \rho^{(\ell)} || \rho_\NATS \big)
   \big\rangle_\ell
   \leq \frac{\theta}{\sqrt{\Sites}} + \theta'. 
\end{align}
The $\Sites$-independent constants $\theta, \theta'$  depend on parameters (the number $c$ of charges, their expectation values, etc.). As the global system grows ($\Sites \to \infty$), the $1 / {\sqrt{\Sites}} \to 0$, so the distance shrinks. 
Hence a physical argument, based on an ensemble in an AMC subspace, complements Jaynes's information-theoretic derivation of $\rho_\NATS$. Furthermore, the AMC subspace enabled later noncommuting-charge work~\cite{NYH_20_Noncommuting, Kranzl_23_Experimental, Murthy_23_Non, Majidy_23_Non}, described later in this section and in Sec.~\ref{sec_new_phys}.

$\rho_\NATS$'s form was justified also with the principle of complete passivity~\cite{Lostaglio_17_Thermodynamic,NYH_16_Microcanonical,Mitsuhashi_22_Characterizing}. A state $\rho$ is \emph{completely passive} if no work can be extracted from $\rho^{\otimes \Sites}$ adiabatically, for any 
$\Sites \in \mathbb{Z}_{\geq 0}$~\cite{Pusz_78_Passive,Lenard_78_Thermodynamical}. 
%
%
Complete passivity extends Carnot's version of the second law of thermodynamics: No engine can extract work from one fixed-temperature heat bath. 
Consider work extraction that evolves a battery's state from $\rho^{(\batt)}$ to $\rho_{\rm f}^{(\batt)}$.
Let $Q_a^{(\batt)}$ denote the battery's $a^\th$ charge.
One can define the (average) work performed on the battery as 
$\Tr (\rho_{\rm f}^{(\batt)} [ \sum_a \mu_a Q_a^{(\batt)} ])
- \Tr (\rho^{(\batt)} [ \sum_a \mu_a Q_a^{(\batt)} ])$. 
Using this definition and a resource theory, one can prove that $\rho_\NATS$ is completely passive~\cite{NYH_16_Microcanonical}. This result resolved the problem spotlighted in~\cite{NYH_18_Beyond}: Noncommuting charges invalidate a resource-theoretic derivation of the thermal state's form.\footnote{
By \emph{invalidate}, we mean that, if the charges fail to commute, the derivation becomes unsound. We use \emph{unsound} in the technical sense of the study of logic: A deductive argument is sound if, provided that the input premises are true, the conclusion is true.}
Yet work is traditionally a change in the energy charge~\cite{Lostaglio_17_Thermodynamic}. 
Indeed, Mitsuhashi \emph{et al.} proved $\rho_\NATS$'s complete passivity while defining work as a change in energy alone (imposing charge conservation on just the extracted-from system)~\cite{Mitsuhashi_22_Characterizing}.
Furthermore, one can define an analog $W_a$ of work for each charge $Q_a$~\cite{Lostaglio_17_Thermodynamic}, as in~\cite{Guryanova_16_Thermodynamics}. 
Consider substituting different $W_a$'s into the definition of complete passivity. Different $W_a$'s will anoint different thermal states as passive, because the $Q_a$'s fail to commute~\cite{Lostaglio_17_Thermodynamic}. 
Hence complete passivity is another thermodynamic principle complicated by noncommuting charges.

The trio~\cite{Guryanova_16_Thermodynamics, Lostaglio_17_Thermodynamic,NYH_16_Microcanonical} galvanized noncommuting-charge activity at the intersection of quantum information and quantum thermodynamics. Several publications elaborated on resource theories~\cite{Gour_18_Quantum,Sparaciari_20_First,Popescu_20_Reference,Khanian_20_Resource,Khanian_20_Quantum}. For example, Gour \emph{et al.} proved more-detailed ``second laws,'' using the mathematical tools of matrix majorization~\cite{Gour_18_Quantum}. These second laws stipulate necessary and sufficient conditions for any quantum state $\sigma_1$ to transform into any state $\sigma_2$. Other literature showed how different batteries could store different $Q_a$'s despite the charges' noncommutation~\cite{Marvian_08_Building,Popescu_20_Reference,Khanian_20_Resource,Bera_19_Quantum}.\footnote{
Batteries are called \emph{reference frames} when used to implement otherwise forbidden transformations. Consider transforming a system-and-environment composite $\Sys \Env$ with unitaries $U$ restricted to conserving each $Q_a^{(\Sys)} + Q_a^{(\Env)}$. What if a desired $U$ breaks the conservation law? A reference frame may supply the charge needed, or accept the charge forfeited, by $\Sys \Env$.}$^,$\footnote{
Reference~\cite{Marvian_08_Building} predates the trio~\cite{Guryanova_16_Thermodynamics, Lostaglio_17_Thermodynamic,NYH_16_Microcanonical}, lacking thermodynamic motivations but having thermodynamic implications.}

The above work, steeped in information theory, is theoretical and abstract. Yet it 
inspired the first experimental demonstration that real-world systems exhibit the thermodynamics of noncommuting charges~\cite{Kranzl_23_Experimental}. Kranzl \textit{et al.} trapped 6--21 ions in a linear Paul chain. Two qubits formed $\mathcal{S}$, and the rest formed $\mathcal{E}$ (Fig.~\ref{fig_Basic_Setup}). 
The global system was prepared in 
$\ket{\psi_0} = \left(\ket{y+}\ket{x+}\ket{z+}\right)^{\otimes N/3}$,
in an AMC subspace.\footnote{
Reference~\cite{NYH_20_Noncommuting} adapted the AMC subspace's definition from the information-theoretic Ref.~\cite{NYH_16_Microcanonical} to many-body physics. In~\cite{NYH_16_Microcanonical}, if $\Sys \Env$ occupies an AMC subspace, measuring any $Q_a^\tot$ has a high probability of yielding a value near $\langle Q_a^\tot \rangle$. Constant-in-$\Sites$ expressions quantify what ``high'' and ``near'' mean. In~\cite{NYH_20_Noncommuting,Kranzl_23_Experimental,Majidy_23_Non}, the probability distribution has a variance of $O( \Sites^r )$, wherein $r \leq 1$.}
The native interactions were long-range Ising couplings
($\vec{\sigma}^\JParen \cdot \vec{\sigma}^\KParen$). However, 
Heisenberg interactions (also long-range) were simulated via Trotterization with rotations.
The dynamics conserved the charges $\sigma_{x,y,z}^{\rm tot}$ globally
while transporting $\sigma_{x,y,z}$ locally.
Quantum state tomography revealed the long-time state $\rho_{\rm f}$ of $\mathcal{S}$. 

Following~\cite{NYH_16_Microcanonical}, Kranzl \textit{et al.}~calculated the state's distance from $\rho_{\rm{NATS}}$, using the relative entropy $D(\rho_{\rm f}||\rho_{\rm NATS})$. They averaged $D(\rho_{\rm f}||\rho_{\rm NATS})$ over the qubit pairs $\mathcal{S}$ that formed the chain, producing $\expval{D(\rho_{\rm f}||\rho_{\rm NATS})}$. Similarly, the authors calculated $\expval{D(\rho_{\rm f}||\rho_{\rm GC})}$ and $\expval{D(\rho_{\rm f}||\rho_{\rm can})}$. 
The NATS predicted the experimental results best: 
$\expval{D(\rho_{\rm f}||\rho_{\rm NATS})} < \expval{D(\rho_{\rm f}||\rho_{\rm GC})} 
< \expval{D(\rho_{\rm f}||\rho_{\rm can})}$. 
Despite the plurality of the conservation laws that decoherence could break, noncommuting charges' effects were experimentally observable; dynamical-decoupling pulse sequences mitigated the noise sufficiently. This experiment, a realization of the proposal~\cite{NYH_20_Noncommuting}, opens the door to experimentally testing the many theoretical results about noncommuting thermodynamic charges, discussed in the next section. Along with trapped ions, feasible platforms include superconducting qudits, ultracold atoms, nitrogen-vacancy centers, and quantum dots~\cite{NYH_20_Noncommuting}.

\section{New physics}
\label{sec_new_phys}

Noncommuting thermodynamic charges engender new physics:
They conflict with the ETH, allow for multiple stationary states, and violate the dichotomy of energy-level statistics. Charges' noncommutation also constrains the global dynamics implementable with local charge-conserving unitaries. Additionally, noncommuting charges decrease thermodynamic entropy production, yet increase average entanglement entropy.

\textit{Conflict with the eigenstate thermalization hypothesis:}
The ETH explains how reversible dynamics thermalize closed quantum many-body systems internally~\cite{Deutsch_91_Quantum, Srednicki_94_Chaos, Rigol_08_Thermalization}.
Consider a system with $N$ DOFs governed by a nondegenerate Hamiltonian $H^\tot$.
Let $\obs$ denote any operator, which we represent as a matrix relative to the energy eigenbasis.
The ETH is an ansatz for the matrix elements' forms.
Suppose that the system begins in a ``microcanonical'' state $|\psi(0)\rangle$---in this context, a pure state with a small $H^\tot$ variance.
The ETH implies thermalization: 
The time-averaged expectation value of $\obs$ approximately equals the canonical expectation value:
\begin{align}\label{eq:time_average}
   \lim_{t\rightarrow \infty} \frac{1}{t} \int_0^t dt'\, 
   \langle \psi(t') |\obs|\psi(t')\rangle 
   = \Tr(\obs \rho_\text{can}) + O(N^{-1}).
\end{align}
Throughout this subsection, big-$O$ notation means ``scales as.''
Noncommuting charges impede three assumptions behind the argument for thermalization: (i) Noncommuting charges prevent microcanonical states from existing in abundance (Sec.~\ref{sec_early_work}). 
(ii) Noncommuting charges force degeneracies on the Hamiltonian (Sec.~\ref{sec_intro}).
(iii) Noncommuting charges lead the Wigner--Eckart theorem to supersede the ETH~\cite{Shankar_94_Principles}.
To review the Wigner--Eckart theorem briefly, we consider
$N$ qubits whose global spin components $S_{x,y,z}^\tot$ are conserved, as in Sec.~\ref{sec_intro}.
Denote by $\{ |\alpha, m\rangle  \}$ the eigenbasis shared by $H^\text{tot}$, 
$( \vec{S}^\tot )^2$, and $S_z^\tot$: If $\hbar = 1$, then
$H^\tot |\alpha, m\rangle = E_\alpha|\alpha,m\rangle$, 
$(\vec{S}^\tot )^2|\alpha,m\rangle = s_\alpha(s_\alpha + 1)|\alpha,m\rangle$, and 
$S_z^\tot |\alpha,m\rangle = m|\alpha,m\rangle$.
The Wigner--Eckart theorem governs spherical tensor operators formed from components $T_q^{(k)}$~\cite{Shankar_94_Principles}. 
The $T_q^{(k)}$'s form a basis for the space of operators defined on the system's Hilbert space. 
For example, consider an atom absorbing a photon (of spin $k=1$), gaining $q=1$ quantum of $z$-type angular momentum. 
$T^{(k=1)}_{q=1}$ represents the photon's effect on the atom's state.
Consider representing $T^{(k)}_q$ as a matrix relative to the energy eigenbasis. That matrix obeys the Wigner--Eckart theorem~\cite{Shankar_94_Principles}: 
\begin{equation}\label{eq:Wigner--Eckart}
    \langle \alpha,m |T_q^{(k)}|\alpha', m'\rangle = \langle s_\alpha,m|s_{\alpha'},m' ;k,q\rangle \langle \alpha ||T^{(k)}||\alpha'\rangle.
\end{equation}
$\langle s_\alpha,m|s_{\alpha'},m' ;k,q\rangle$ denotes a Clebsch-Gordan coefficient, a conversion factor between the product state 
$| s_{\alpha'}, m'; k, q \rangle
\equiv | s_{\alpha'}, m' \rangle  \ket{k, q}$ 
and the total-spin eigenstate $| s_\alpha, m \rangle$. 
$\langle \alpha ||T^{(k)}||\alpha'\rangle$ is a reduced matrix element---the part of $\langle \alpha, m | T^{(k)}_q | \alpha', m' \rangle$ that does not depend on magnetic spin quantum numbers. The theorem~\eqref{eq:Wigner--Eckart} conflicts with the ETH, an ansatz for the left-hand side. First, the two constraints are fundamentally different; the Wigner--Eckart theorem concerns SU(2) symmetry,  while the ETH concerns randomness.
Second, the ETH states that off-diagonal elements $\langle \alpha,m |T_q^{(k)}|\alpha', m'\rangle$ are exponentially small in $N$. 
The Wigner--Eckart theorem implies that these elements may be $O(1)$.

Reference~\cite{Murthy_23_Non} therefore posited a \emph{non-Abelian ETH.} 
This ansatz depends on the average energy 
$\mathscr{E}  \coloneqq  \frac{1}{2}(E_\alpha + E_{\alpha'})$, energy difference 
$\omega  \coloneqq  E_\alpha - E_{\alpha'}$, average spin quantum number 
$\mathscr{S}  \coloneqq  \frac{1}{2}(s_\alpha + s_{\alpha'})$, and difference 
$\nu  \coloneqq  s_\alpha - s_{\alpha'}$. 
Denote by $S_\text{th}(\mathscr{E},\mathscr{S})$ the thermodynamic entropy at energy $\mathscr{E}$ and spin $\mathscr{S}$.
The observable $T_q^{(k)}$ and Hamiltonian $H^\text{tot}$ satisfy the non-Abelian ETH if, for smooth real functions $\mathcal{T}^{(k)}(\mathscr{E},\mathscr{S})$ and $f_\nu^{(k)}(\mathscr{E},\mathscr{S},\omega)$,
\begin{equation}\label{eq:non-Abelian_ETH}
    \begin{split}
    \langle \alpha ||T^{(k)}||\alpha'\rangle
    = & \mathcal{T}^{(k)} (\mathscr{E}, \mathscr{S}) \, \delta_{\alpha,\alpha'} \\
    &+ e^{-S_\text{th} (\mathscr{E},\mathscr{S})/2} \,
    f_\nu^{(k)}(\mathscr{E},\mathscr{S}, \omega) \, R_{\alpha,\alpha'}.
    \end{split}
\end{equation}
$\mathcal{T}^{(k)}$ parallels a microcanonical average in the conventional ETH.
$R_{\alpha, \alpha'}$ resembles a normalized $O(1)$ random variable~\cite{Foini_19_Eigenstate, Pappalardi_22_Eigenstate, Wang_22_Eigenstate}.
The matrix element \eqref{eq:Wigner--Eckart} deviates from the ordinary ETH through $\mathscr{S}$-dependent functions and a Clebsch--Gordan coefficient. 
Equation~\eqref{eq:non-Abelian_ETH}
has withstood preliminary numerical checks with a Heisenberg Hamiltonian on a 2-dimensional qubit lattice~\cite{Noh_22_Eigenstate}. More testing is needed, however.

The non-Abelian ETH predicts thermalization to the usual extent in some, but not all, contexts.
Consider preparing the system in a state $|\psi(0)\rangle$ in an AMC subspace (Sec.~\ref{sec_early_work}). 
Suppose that $|\psi(0)\rangle$ has an extensive magnetization along an axis that we call $\hat{z}$: $\langle \psi(0)| S_z^\tot |\psi(0) \rangle = O(\Sites)$. According to the non-Abelian ETH,
\begin{equation}
\begin{split}
    &\lim_{t\rightarrow \infty} \frac{1}{t} \int_0^{t} dt'\, 
    \langle \psi(t')|T_q^{(k)}|\psi(t')\rangle \\
    &\quad = \Tr\left(T_q^{(k)} \rho_\text{NATS}\right) + O\left( \Sites^{-1} \right).
    \end{split}
\end{equation} 
However, if $\langle \psi(0) | S_z^\tot | \psi(0) \rangle = 0$, the correction can become $O( \Sites^{-1/2} )$---polynomially larger.
This result relies on an assumption argued to be physically reasonable: The smooth function $\mathcal{T}^{(k)}$ in~\eqref{eq:non-Abelian_ETH} can contain a nonzero term of $O(s_\alpha / \Sites )$.
The unusually large, analytically expected correction constitutes further evidence that charges' noncommutation can alter thermalization. 

The non-Abelian ETH invites further studies---first, numerical and experimental tests. Second, from the non-Abelian ETH's final term, one can infer about fluctuations around the time average~\eqref{eq:time_average}. 
Third, the time required for equilibration merits study: In quantum many-body physics, most thermalization-blocking mechanisms slow thermalization in time (Sec.~\ref{sec_adjacent}). In contrast, the $O(N^{-1/2})$ ``slows'' the time-averaged expectation value in $\Sites$, in the approach to a thermal expectation value.
How long does 
$\langle \psi(t')|T_q^{(k)}|\psi(t')\rangle$ 
take to reach its long-time value?
Fourth, using the ETH, Strasberg \textit{et al.}~demonstrated how classicality, Markovianity, and local detailed balance emerge from the pure-state dynamics of a system with commuting charges~\cite{Strasberg_22_Classicality}. Their results might be extended to noncommuting-charge systems, with the non-Abelian ETH and AMC subspaces.
Finally, one might store information in a system that thermalizes---forgets its initial conditions---less than usual. This memory-storage opportunity resonates with the work that we explain now.

\emph{Stationary-state multiplicity:} Consider an open quantum system with a Liouvilian superoperator $\mathscr{L}$. $\rho_{\rm stat}$ is a stationary state of $\mathscr{L}$ if $\mathscr{L}\rho_{\rm stat} = 0$. 
Consider an arbitrary basis for the stationary subspace.
One might associate the $j^\th$ basis element,
$\rho_{\rm stat}^\JParen$, with a classical alphabet's $j^\th$ letter, $L_j$: 
$\rho_{\rm stat}^\JParen  \leftrightarrow  L_j$.
To encode $L_j$ in the system, one would prepare any state that thermalizes to $\rho_{\rm stat}^\JParen$.
The larger the stationary subspace,
the more classical information the system may store.
Denote by $n_{\rm stat}$ the stationary subspace's dimensionality. 
Zhang \textit{et al.}~derive a lower bound $n_{\rm NC}$ on $n_{\rm stat}$ for a system with noncommuting charges~\cite{Zhang_20_Stationary}: $n_{\rm NC} = \sum_j \mathcal{D}_j^2$, wherein $\mathcal{D}_j$ denotes the symmetry group's $j^\th$ irreducible representation. For example, $n_{\rm NC} \sim N^3$ for a Heisenberg model (Sec.~\ref{sec_intro}) coupled to an environment that conserves the system-of-interest charges 
$\sigma_{x,y,z}^{(\Sys)}$.
If the charges commute, the lower bound $n_{\rm C}$ scales as the number of simultaneous eigenspaces shared by all the $Q_a^{(\Sys)}$'s. Since $n_{\rm NC}$ and $n_{\rm C}$ scale differently, noncommuting charges could alter the stationary subspace's dimensionality and so the amount of information storable.

\textit{Hybrid energy-level statistics:}
Energy-level statistics diagnose quantum chaos and integrability~\cite{DAlessio_16_From}.
Denote by $\omega = E_{j+1} - E_j$ the spacing between consecutive eigenenergies of a many-body Hamiltonian.
Any given spacing (near the spectrum's center) has a probability density $P(\omega)$ of having the size $\omega$. 
A Poissonian\footnote{
This $P(\omega)$ is the Poisson distribution whose average-rate-of-occurrence parameter vanishes. The reason is, $P(\omega) \, d\omega$ equals the probability that zero eigenenergies lie in a width-$d\omega$ interval.} 
$P(\omega)$ diagnoses integrability ~\cite[Sec.~2.3]{DAlessio_16_From}; and a Wigner--Dyson distribution, 
$A_\gamma  \,  \omega^\gamma  \exp(-B_\gamma \omega^2)$, chaos~\cite{Wigner_Characteristic_1955,Dyson_62_Statistical}.
The parameter $\gamma \in \{1,2,4\}$ 
depends on the Hamiltonian's time-reversal and rotational symmetries. 
Normalization and the mean $\omega$ determine the coefficients 
$A_\gamma$ and $B_\gamma$.
Noncommuting charges defy the Poisson-vs.-Wigner--Dyson dichotomy as commuting charges cannot:
The charges generate a non-Abelian algebra, which has multidimensional irreducible representations, which force degeneracies on the Hamiltonian.
These representations induce statistics that interpolate between the two distributions~\cite{Rosenzweig_60_Repulsion, Noh_22_Eigenstate, Giraud_22_Probing}.
Noh observed such statistics numerically, using a 2D Heisenberg model~\cite{Noh_22_Eigenstate}.

\textit{Constraints on charge-conserving dynamics:}
Natural interactions are spatially local, 
motivating a quantum-computational result: Every $\Sites$-qubit unitary decomposes into gates on pairs of qubits~\cite{DiVincenzo_95_Two, Lloyd_95_Almost, Deutsch_97_Universality}.
Can charge-conserving local unitaries effect every charge-conserving global unitary $U$?
Perhaps surprisingly, the answer is no~\cite{Marvian_22_Restrictions}.
Using Lie theory, Marvian proved that
locality-constrained charge-conserving unitaries
cannot even approximate $U$'s.
The reason is, the two types of unitaries form Lie groups of different dimensions
\cite{Marvian_22_Restrictions,Marvian_21_Qudit,Marvian_22_Rotationally, Marvian_23_Non}.
Moreover, noncommuting charges impose four types of constraints on the implementable global unitaries, whereas commuting charges impose only two~\cite{Marvian_23_Non}. 
The extra constraints may restrict chaos, which enables thermalization. 
Hence Marvian's result suggests that noncommuting charges may restrict thermalization.

\begin{figure}
    \centering    
    \includegraphics[width=0.9\columnwidth]{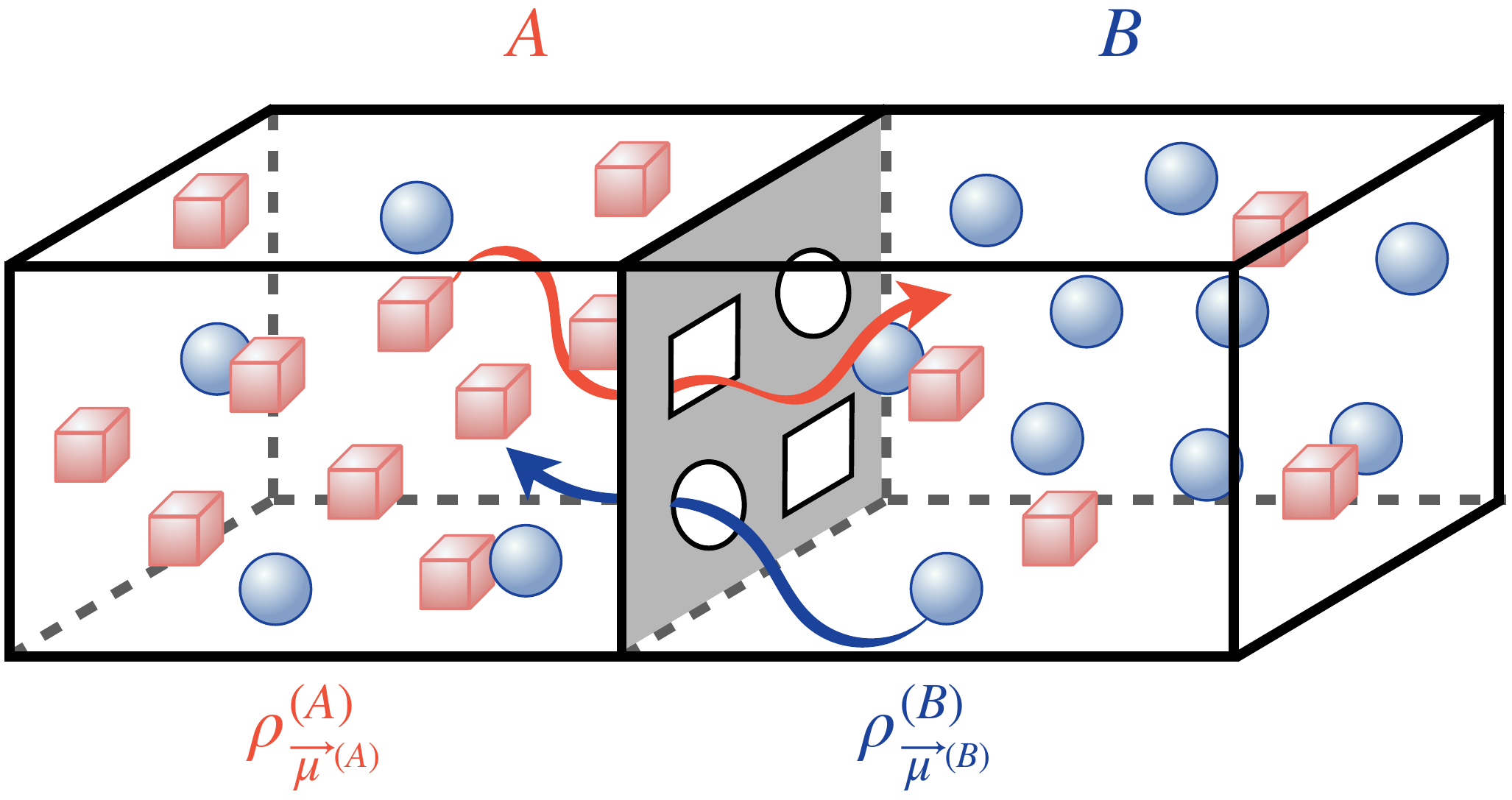}
    \caption{\caphead{Two thermal reservoirs exchange charges, producing entropy:} 
    Blue spheres represent charges of one type, and red cubes represent charges of another.}
    \label{fig_transport}
\end{figure}

\textit{Decreased thermodynamic-entropy production:}
Noncommuting charges reduce entropy production, 
which quantifies irreversibility~\cite{Manzano_22_Non}.
Throughout this paragraph and the next, \emph{entropy} refers to thermodynamic entropy,
not entanglement entropy.
Let systems $X = A, B$ have charges $Q_a^{(X)}$ that might or might not commute. (In no other subsection of this section do $Q_a^{(X)}$'s denote possibly commuting charges.) Each system begins in a GGE 
$\rho^{(X)}_{\vec{\mu}^{(X)}} \propto \exp \left(\sum_a \mu_a^{(X)} Q_a^{(X)}\right)$, wherein $\vec{\mu}^{(X)} 
= (\mu_0^{(X)}, \mu_1^{(X)}, ...)$~\cite{Manzano_18_Squeezed} 
(Fig. \ref{fig_transport}). 
Hence $AB$ begins in
$\rho(0) \coloneqq \rho^{(A)}_{\vec{\mu}^{(A)}}  \otimes 
\rho^{(B)}_{\vec{\mu}^{(B)}}$.
Let us specialize to the linear-response regime: 
$\vec{\mu}^{(A)} \approx \vec{\mu}^{(B)}$.
A charge-conserving unitary $U$ can shuttle charges between the systems, producing entropy. Noncommuting charges decrease the entropy production~\cite{Manzano_22_Non}:
Denote by $\delta \mu_a  \coloneqq  \mu_a^{(A)} - \mu_a^{(B)}$ the difference between the systems' $a$-type chemical potentials. 
System $A$'s $a$-type charge changes, in the Heisenberg picture, by 
$U^\dag Q_a^{(A)} U - Q_a^{(A)}$. 
(Tensored-on $\id$'s are implicit where necessary to make operators act on the appropriate Hilbert spaces.)
We combine the foregoing two quantities into 
$\epo \coloneqq \sum_a \delta \mu_a \big( U^\dagger Q_a^{(A)} U - Q_a^{(A)} \big)$. 
Taking the expectation value of $\epo$ in the initial state, we obtain the net entropy production,
$\Sigma = \Tr  \LParen  \tilde{\Sigma} \,  \rho(0)  \RParen$.
The initial state lies near the fixed point 
$\pi \coloneqq \rho^{(A)}_{\vec{\mu}^{(A)}}\otimes \rho^{(B)}_{\vec{\mu}^{(A)}}$ 
of U, by the linear-response assumption.\footnote{
In $\pi$, unlike in $\rho(0)$, $A$ and $B$ have the same lists of chemical potentials.}
$\pi$ and $\epo$ have a Wigner--Yanase--Dyson (WYD) skew information $I_y(\pi,\epo) \coloneqq -\frac{1}{2}\Tr  \big(  [\pi^y,\epo][\pi^{1-y},\epo]  \big)$, 
parameterized by $y \in (0, 1)$. 
$I_y(\pi,\epo)$ quantifies the coherence that $\epo$ has relative to $\pi$'s eigenbasis.
The WYD skew information contributes to the entropy production:
$\Sigma = \frac{1}{2}\var(\epo) - \frac{1}{2} \int_0^1 \text{d}y\, I_y(\pi,\epo)$~\cite{Manzano_22_Non}.
 $I_y$ is always $\geq 0$ and is positive if and only if the charges fail to commute.
 Therefore, noncommuting charges lower $\Sigma$. Since entropy production accompanies thermalization, noncommutation may inhibit thermalization. 
 
Two extensions support~\cite{Manzano_22_Non}. First, Shahidani numerically simulated an optomechanical system interacting with a squeezed thermal bath~\cite{Shahidani_22_Thermodynamic}. Second, Upadhyaya \emph{et al.} progressed beyond the linear-response regime~\cite{Upadhyaya_23_What}. Even there, charges' noncommutation can decrease entropy production.

How~\cite{Manzano_22_Non} impacts engine efficiencies remains an open question. Efficiencies suffer from waste and so may benefit from noncommuting charges' lowering of $\Sigma$. However, charges' noncommutation can decrease the amount $W_a$ of $a$-charge (analogous to work) extractable by an engine~\cite{Ito_18_Optimal}. If using finite-size NATS baths, an engine can extract 
$W_a \lesssim W_a \rvert_{\text{infinite-size bath}} - O (C / \xi)$. 
$C$ captures the correlations between the charges in $\rho_\NATS$; if the charges commute, $C = 0$.
$\xi$, encoding the baths' sizes, diverges in the infinite-bath limit. 
Extractable work, here lowered by charges' noncommutation, tends to trade off with efficiency. A precise relationship between the two figures of merit, in the presence of noncommuting charges, merits calculation.

\textit{Increased average entanglement entropy:}
Quantum many-body thermalization entails entanglement, which charges' noncommutation can enhance 
\cite{Majidy_23_Non}. 
Consider initializing an isolated $N$-qubit system in a pure state. Divide the system into subsystems $A$ and $B$ with sizes $N_A$ and $N_B$. 
$A$ and $B$ can share entanglement quantified with the \emph{entanglement entropy}, the von Neumann entropy of the reduced state $\rho_A$ of $A$: $S_{AB}  \coloneqq  S_\text{vN}(\rho_A)$.
Plotting $S_{AB}$ against $N_A$ yields the \emph{Page curve}~\cite{Page_93_Average}.

\begin{figure}
    \centering
    \includegraphics[width=0.9\columnwidth]{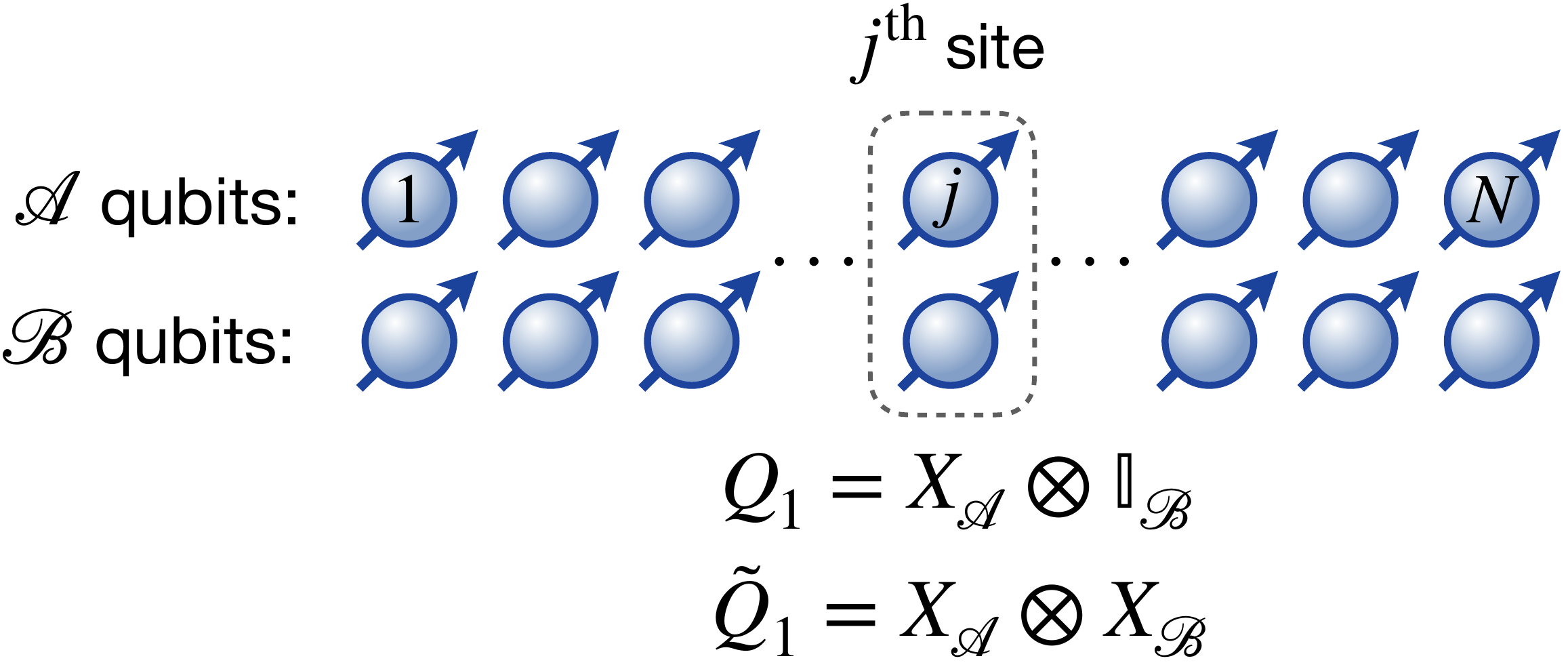}
    \caption{ \caphead{Analogous noncommuting-charge and commuting-charge models:} 
    Each model consists of $\Sites$ sites, each formed from $\mathcal{A}$ and $\mathcal{B}$ qubits. 
    Local observables distinguish the models. The noncommuting observables are $Q_1 = X_\mathcal{A} \otimes \mathbbm{1}_\mathcal{B}$, $Q_2 = Y_\mathcal{A}\otimes \mathbbm{1}_\mathcal{B}$, and $Q_3 = Z_\mathcal{A} \otimes \mathbbm{1}_\mathcal{B}$. The commuting observables are $\tilde{Q}_1 = X_\mathcal{A} \otimes X_\mathcal{B}$, $\tilde{Q}_2 = Y_\mathcal{A}\otimes Y_\mathcal{B}$, and $\tilde{Q}_3 = Z_\mathcal{A} \otimes Z_\mathcal{B}$.
    The local charges, summed across the chain, form global observables that define the (approximate) microcanonical subspaces.
    (Adapted from~\cite{Majidy_23_Non}.)}
    \label{fig_Page_curve_models}
\end{figure}

Majidy \textit{et al.} computed Page curves for two comparable $N$-qubit models, one with commuting charges only and the other with noncommuting charges~\cite{Majidy_23_Non}. 
Figure~\ref{fig_Page_curve_models} describes the models.
They involve no dynamics, so the observables of interest are not technically charges. However, the authors modeled states resulting from chaotic dynamics:
Haar-random states were sampled from (approximate, when necessary~\cite{NYH_16_Microcanonical, NYH_20_Noncommuting,
Kranzl_23_Experimental}) microcanonical subspaces.
Page curves were estimated numerically via exact diagonalization and analytically via large-$N$ approximations of combinatorics.
The noncommuting-charge Haar-averaged Page curves lay above the commuting-charge analogs. 
The difference was of $O\big(\frac{N_A^2}{N^2 N_B}\big)$, in the simplest comparison.
A possible reason centers on the least entangled basis for each model's subspace. 
If the local charges $\tilde{Q}_a$ commute, they share an eigenbasis. Hence the global commuting $\tilde{Q}_a^\tot$'s share a tensor-product eigenbasis for the subspace. If the local charges fail to commute, this argument breaks, and the subspace's least entangled basis is entangled. One might therefore expect more entanglement of the noncommuting-charge model on average across the subspace.

Other works concern non-Abelian symmetries' effects on entanglement entropy but focus less on what charges' noncommutation changes.
For example, a non-Abelian symmetry raises the entanglement in Wess--Zumino--Witten models, $(1+1)$-dimensional conformal field theories with Lie-group symmetries~\cite{Calabrese_21_Symmetry, Goldstein_18_Symmetry}.
Second, holographic calculations highlight another correction that non-Abelian symmetries introduce into entanglement entropy~\cite{Zhao_21_Symmetry, Zhao_22_Charged}.\footnote{
This correction appears to be negative. However,~\cite{Zhao_21_Symmetry, Zhao_22_Charged} concern symmetry-resolved Page curves, in contrast with the conventional Page curves of~\cite{Majidy_23_Non}. A symmetry-resolved Page curve models the entanglement, averaged over time, of a system whose charges move only within $A$ and within $B$, not between the subsystems. Conventional Page curves model less-restricted thermalization.}
Third, algebraic-quantum-field-theory calculations agree that non-Abelian symmetries raise Page curves~\cite{Bianchi_19_Typical}.

These works suggest several research opportunities.
The entanglement entropy's increase merits checking with more comparable models \`{a} la Fig.~\ref{fig_Page_curve_models}.
Additionally, one might adjust the Page-curve calculations following
Marvian's revelation that local charge-conserving unitaries constrain global $U$'s tightly~\cite{Marvian_22_Restrictions}.
Under locality constraints, the Haar distribution may model chaotic dynamics inaccurately.
Finally, the entanglement entropy of~\cite{Majidy_23_Non,Calabrese_21_Symmetry, Goldstein_18_Symmetry,Zhao_21_Symmetry, Zhao_22_Charged,Bianchi_19_Typical} differs from the thermodynamic entropy of~\cite{Manzano_22_Non}. Still, noncommuting charges' increase of the former conceptually conflicts, somewhat, with their decrease of the latter.

\section{Adjacent work in other fields}
\label{sec_adjacent}

Noncommuting charges have arisen in other thermodynamics-related contexts, usually under the guise of non-Abelian symmetries. We discuss four contexts: First, integrable systems have more charges, which might fail to commute, than the thermodynamic systems of Sections~\ref{sec_intro}--\ref{sec_new_phys}. Second, noncommuting charges destabilize a thermalization-avoidant phase of quantum matter: many-body localization (MBL). Third, gauge theories naturally have non-Abelian symmetries, albeit local ones. Fourth, hydrodynamics describes the flow of some non-Abelian charges, including in heavy-ion collisions. 

Three more connections, we mention briefly: Noncommuting charges may impact quantum information scrambling~\cite{Agarwal_22_Emergent}. Also, $\rho_\NATS$ helps explain phase transitions~\cite{Corps_22_Dynamical,Corps_22_Theory,Corps_23_General}. Finally, the resource theory of asymmetry quantifies a state's noninvariance under operation by a symmetry group's elements~\cite{Gour_08_Resource}. The group may be non-Abelian, so the theory's mathematical tools merit application to noncommuting thermodynamic charges.

\emph{Conventional integrable systems:} Garden-variety integrable systems fail to thermalize. Each has extensively many nontrivial charges, which constrain the dynamics significantly. In contrast, our thermodynamic setup (Sec.~\ref{sec_intro}) entails a number of charges, $c \ll \Sites$, much less than the number of DOFs. 
Integrable and near-integrable systems with noncommuting charges have been studied.
Models include the GGE equal to $\rho_\NATS$~\eqref{eq_NATS}. Several works assess how the GGE's accuracy grows with the number of $Q_a$'s included in the state~\cite{Fagotti_14_Conservation,Fagotti_17_Charges,Fukai_20_Noncommutative,Corps_23_General}. Outside of GGE studies, an integrable system can exhibit anomalous transport---diffusion with a diffusion constant $D \sim \sqrt{\Sites}$~\cite{Znidaric_11_Spin,Ljubotina_17_Spin}. 
The system studied in~\cite{Znidaric_11_Spin} is the one-dimensional nearest-neighbor Heisenberg model~\eqref{eq_Heis}.
The anomalous transport can be explained with Bethe-ansatz calculations and
the $z$-axis magnetization~\cite{Gopalakrishnan_19_Kinetic,De_21_Stability,Ilievski_21_Superuniversality}. Can all three noncommuting charges ($\sigma_{x,y,z}$) predict anomalous transport alternatively? Another open question follows from increasingly breaking a noncommuting-charge system's integrability~\cite{Mierzejewski_22_Multiple}. How do the system's behaviors transmute into thermodynamic behaviors described in this Perspective?

\emph{Many-body localization:} Disordered, interacting quantum systems exhibit MBL. Examples include a qubit chain subject to the disordered Heisenberg Hamiltonian
$H^\tot_\MBL = J \sum_{j=1}^\Sites  \big( \vec{\sigma}^\JParen  \cdot  \vec{\sigma}^{(j+1)}
+  h_j  \sigma_z^\JParen  \big)$.
The disorder term, $\sum_{j=1}^\Sites  h_j  \sigma_z^\JParen$,
acts as an external field whose magnitude $h_j$ varies randomly across sites.
Denote by $h$ the disorder's standard deviation.
If the disorder is much stronger than the interaction, $h \gg J$,
the system localizes: Imagine measuring each qubit's $\sigma_z$. The qubits approximately maintain the measured configuration long afterward. This behavior contrasts with how thermalizing systems, such as classical gases, change configurations quickly. Hence MBL resists thermalization for long times. The reason is, the Hamiltonian decomposes as a linear combination of quasilocal DOFs~\cite{Abanin_19_Colloquium}.

Noncommuting charges destabilize MBL~\cite{Potter_16_Symmetry}. Consider forcing a non-Abelian symmetry on $H^\tot_\MBL$. The resulting Hamiltonian, $H^\tot_{\MBL'}$, will have degeneracies, by Schur's lemma (App.~\ref{sec_Schur}). So will the quasilocal DOFs, which can therefore become ``excited'' at no energy cost. Consider adding to $H^\tot_{\MBL'}$ an infinitesimal field that violates the symmetry. The resulting Hamiltonian, $H^\tot_{\MBL''}$, can map $H^\tot_{\MBL'}$ eigenstates $\ket{\psi}$ to same-energy eigenstates $\ket{\tilde{\psi}}$: 
$\bra{\tilde{\psi}}  H^\tot_{\MBL''}  \ket{\psi}  \neq  0$. Two such eigenstates can be zero-energy ``excited'' states of neighboring quasilocal DOFs. Hence $H^\tot_{\MBL''}$ can transport zero-energy ``excitations'' between quasilocal DOFs---across the system. Such transport is inconsistent with MBL. Therefore, non-Abelian symmetries promote a thermalizing behavior.

\emph{Gauge theories:} Classical electrodynamics exemplifies a gauge theory, a model that contains more DOFs than does the physical system modeled~\cite{Peskin_95_Introduction}. Choosing a gauge eliminates the extra DOFs. The transformations between gauges form a Lie group $\mathcal{G}$. The elements $U \in \mathcal{G}$ preserve the theory's action, $\mathscr{S}$. That is, $U : \mathscr{S} \mapsto \mathscr{S}$. Gauge theories model elementary-particle physics and condensed matter, both of whose $\mathcal{G}$'s can be non-Abelian. For example, quantum chromodynamics, which describes the strong force, has SU(3) symmetry. Hence elementary-particle physics should realize noncommuting-charge thermodynamics naturally; $\rho_\NATS$ might be observable in high-energy and nuclear systems~\cite{Mueller_22_Thermalization}. Granted, confinement prevents quantum-chromodynamic systems from having nonzero $\langle Q_a^\tot \rangle$'s. Still, noncommuting charges raise the average entanglement entropy in spin systems with $\langle Q_a^\tot \rangle = 0$ $\forall a$~\cite{Majidy_23_Non}. Furthermore, subsystems $j$ can contain positive and negative charges, $\langle Q_a^\JParen \rangle \neq 0$, that can undergo dynamics~\cite{NYH_22_How}. Gauge symmetries are local, though, unlike the global symmetries covered in this Perspective. The contrast raises the question of how much noncommuting-charge quantum thermodynamics ports over into gauge theories.

\emph{Hydrodynamics and heavy-ion collisions:} Hydrodynamics models long-wavelength properties of fluids that are in equilibrium locally~\cite{Glorioso_18_Lectures}. The theory describes condensed matter, certain stages of heavy-ion collisions~\cite{Baier_06_Dissipative}, and more. Noncommuting charges can flow similarly to the more-often studied energy and particles~\cite{Glorioso_21_Hydrodynamics}. A few effects of charges' noncommutation have been isolated; examples include non-Abelian contributions to conductivity~\cite{Torabian_09_Holographic} and entropy currents~\cite{Hoyos_14_Odd}. Also, non-Abelian symmetries can shorten charge-neutralization times in heavy-ion collisions~\cite{Elze_89_Quark}. More such effects may be discoverable. Also, quantum thermodynamics might assist with longstanding questions about heavy-ion collisions: by what process the system thermalizes, how the initial state should be modeled, whether the (non-)Abelian ETH explains the thermalization~\cite{Yao_23_SU2,Ebner_23_Eigenstate}, and why the thermalization time is so short~\cite{Berges_21_QCD}.

\section{Outlook}
\label{sec_outlook}

Noncommuting charges pose a quantum challenge to expectations about thermodynamics. Charges' noncommutation impedes derivations of the thermal state's form~\cite{NYH_18_Beyond,NYH_16_Microcanonical}, eliminates microcanonical subspaces~\cite{NYH_16_Microcanonical}, and decreases thermodynamic-entropy production~\cite{Manzano_22_Non}.
Dynamically, noncommuting charges limit the global evolutions implementable with local interactions more than commuting charges do~\cite{Marvian_23_Non}.
In many-body physics, noncommuting charges underlie steady-state degeneracies that might be employed to store information~\cite{Zhang_20_Stationary}. Noncommuting charges also conflict with the ETH~\cite{Murthy_23_Non} and can increase average entanglement entropy~\cite{Majidy_23_Non,Calabrese_21_Symmetry, Goldstein_18_Symmetry,Bianchi_19_Typical}. 
These discoveries suggest rich research opportunities, of which we detail five.

First, the predictions merit experimental testing. The first test of noncommuting-charge thermodynamics was performed with trapped ions~\cite{Kranzl_23_Experimental}. Other potential platforms include superconducting qubits, quantum dots, ultracold atoms, quantum optics, and optomechanics~\cite{NYH_20_Noncommuting,NYH_22_How,Manzano_22_Non}.

Second, existing results present a conceptual puzzle. Evidence suggests that noncommuting charges hinder thermalization to an extent: They invalidate derivations of the thermal state's form~\cite{NYH_18_Beyond,NYH_16_Microcanonical}, decrease thermodynamic-entropy production~\cite{Manzano_22_Non}, clash with the ETH~\cite{Murthy_23_Non}, and uniquely restrict the global unitaries implementable via local interactions~\cite{Marvian_23_Non}. Other evidence, however, suggests that noncommuting charges enhance thermalization: They destabilize MBL~\cite{Potter_16_Symmetry} and increase average entanglement entropy~\cite{Majidy_23_Non}. These results do not conflict with each other technically, stemming from different setups. Yet the results clash conceptually, suggesting that charges' noncommutation hinders thermalization in some ways and enhances it in others. Does the hindrance or enhancement win out overall?
Reconciling these results presents a challenge.

Third, to what extent can classical mechanics reproduce noncommuting-charge thermodynamics? Observables' noncommutation underlies quintessentially quantum phenomena including uncertainty relations, measurement disturbance, and the Einstein--Podolsky--Rosen paradox. Yet classical mechanics features quantities that fail to commute with each other---for example, rotations about different axes. How \emph{nonclassical} is noncommuting-charge thermodynamics (achievable only outside of classical physics), beyond being merely \emph{quantum} (achievable within quantum physics)?

Fourth, every chaotic or thermodynamic phenomenon merits re-examination. To what extent does it change under dynamics that conserve noncommuting charges? Example phenomena include diffusion coefficients, transport relations, thermalization times, monitored circuits~\cite{Majidy_23_Critical}, out-of-time-ordered correlators~\cite{Swingle_18_Unscrambling}, operator spreading~\cite{Khemani_18_Operator}, frame potentials~\cite{Roberts_17_Chaos}, and quantum-complexity growth~\cite{Brown_18_Second}.

Finally, noncommuting-charge thermodynamics merits bridging to similar topics in neighboring fields. Non-Abelian gauge theories, non-Abelian hydrodynamics, GGE studies, and dynamical phase transitions overlap with noncommuting thermodynamic charges. To what extent can these areas inform each other? Do gauge theories realize noncommuting thermodynamic charges naturally?

For decades, conserved thermodynamic quantities were assumed implicitly to commute with each other. Noncommutation, however, is a trademark of quantum theory. The identification and elimination of the assumption, though initiated where quantum information theory meets quantum thermodynamics, have potential ramifications across quantum many-body physics.

%
%
\begin{acknowledgments}
This work received support from the John Templeton Foundation (award no. 62422)
and the National Science Foundation (QLCI grant OMA-2120757). 
The opinions expressed in this publication are those of the authors and do not necessarily reflect the views of the John Templeton Foundation or UMD.
S.M. received support from the Vanier C.G.S.
T.U. acknowledges the support of the Joint Center for Quantum Information and Computer Science through the Lanczos Fellowship, as well as the Natural Sciences and Engineering Research Council of Canada (NSERC), through the Doctoral Postgraduate Scholarship.
N.Y.H. thanks the Institut Pascal for its hospitality during the formation of this paper, Maurizio~Fagotti for discussions about GGEs, Niklas~Mueller for discussions about lattice gauge theories and hydrodynamics, and Christopher~D.~White for MBL discussions.
\end{acknowledgments}

\section{Competing interests}

The authors declare no competing interests.

\begin{appendices}


\renewcommand{\thesection}{\Alph{section}}
\renewcommand{\thesubsection}{\Alph{section} \arabic{subsection}}
\renewcommand{\thesubsubsection}{\Alph{section} \arabic{subsection} \roman{subsubsection}}

\makeatletter\@addtoreset{equation}{section}
\def\theequation{\thesection\arabic{equation}}

\section{Schur's lemma implies degeneracy of Hamiltonians that have non-Abelian symmetries}
\label{sec_Schur}


Consider a quantum system associated with a Hilbert space $\mathcal{H}$. Let $H^\tot$ denote the Hamiltonian, which has a symmetry associated with a Lie group $\mathcal{G}$. Let $k$ label the irreducible representations (irreps) of $\mathcal{G}$. For example, $\mathcal{G} =$ SU(2) has irreps labeled by $s = 0, 1/2, 1, \ldots$ A corollary of Schur's lemma states that~\cite[p.~9]{Gour_08_Resource} 
\begin{align}
   \mathcal{H}
   & = \bigoplus_k  \mathcal{H}_k  \otimes  \mathcal{M}_k .
\end{align}
$\mathcal{H}_k$ denotes a representation space. The multiplicity space $\mathcal{M}_k$ has a dimensionality equal to irrep $k$'s dimension. 

Correspondingly, $H^\tot$ decomposes as~\cite[Sec.~3.2.3]{Fulton_04_Representation}
\begin{align}
   H^\tot
   & = \bigoplus_k  E_k  \id_k .
\end{align}
$E_k$ denotes an eigenenergy. $\id_k$ denotes the identity operator acting on the multiplicity space $\mathcal{M}_k$. If $\mathcal{G}$ is non-Abelian, then some of the irreps $k$ have dimensionalities $>1$. Consequently, some $\id_k$'s act on multidimensional subspaces $\mathcal{M}_k$. Those $\mathcal{M}_k$'s are degenerate eigenspaces of $H^\tot$.

\end{appendices}

\onecolumngrid

%
%
\bibliographystyle{ieeetr}
\bibliography{noncommQ_refs}

\end{document}